\documentclass[3p,twocolumn]{elsarticle}

\usepackage{ecrc}

\volume{00}

\firstpage{1}

\journalname{Nuclear and Particle Physics Proceedings}

\runauth{G. Giacinti \& J. G. Kirk}

\jid{nppp}

\jnltitlelogo{Nuclear and Particle Physics Proceedings}

\usepackage{amssymb}

\usepackage{rotating}

\begin{document}

\begin{frontmatter}



\dochead{}

\title{Cosmic-Ray Anisotropy and the Local Interstellar Turbulence}


\author{Gwenael~Giacinti}
\author{John~G.~Kirk}

\address{Max-Planck-Institut f\"ur Kernphysik, Postfach 103980, 69029 Heidelberg, Germany}

\begin{abstract}
  We study the role of local interstellar turbulence in shaping the
  large-scale anisotropy in the arrival directions of TeV--PeV
  cosmic-rays (CRs) on the sky. Assuming pitch-angle diffusion of CRs
  in a magnetic flux tube containing the Earth, we
  compute the CR anisotropy for Goldreich-Sridhar turbulence, and for
  isotropic fast modes.  The narrow deficits in the $400$\,TeV and
  $2\,$PeV data sets of IceTop can be fitted for some parameters of
  the turbulence.  The data also rule out a part of the parameter
  space. The shape of the CR anisotropy may be used as a local probe
  of the still poorly known properties of the interstellar turbulence
  and of CR transport.
\end{abstract}


\end{frontmatter}

\section{Introduction}
\label{Introduction}

In this paper we report on investigations of how the {\em shape} of
the large-scale (LS) anisotropy of TeV--PeV cosmic-rays, (i.e.,
excluding features whose angular sizes are smaller than a few tens of
degrees) depends on the properties of the interstellar turbulence and
cosmic ray (CR) transport within $\sim 10$\,pc from
Earth~\cite{Giacinti:2016tld}. Up until now, most studies of the
large-scale CR anisotropy (CRA) have focussed on the {\em direction}
and {\em amplitude} of the {\em dipole}, and in particular its
relation to local sources of CRs (for a recent study see
e.g.~\cite{Ahlers:2016njd}). The direction of the CRA 
is observed to be compatible with that of the local interstellar
magnetic field, as deduced from the IBEX ribbon and from polarization
of starlight from stars within ten to a few tens of pc from
Earth~\cite{Schwadron2014,Frisch:2012zj,Frisch:2015hfa}. However, the
LS~CRA is not well described by a dipole~\cite{Aartsen:2012ma}. A few
earlier studies have modeled phenomenologically the LS~CRA as the sum
of a dipole and higher order multipoles (e.g.~\cite{Zhang:2014dsu}),
but the study we present here is, to our knowledge, the first
quantitative description linking it to the power-spectrum and other
parameters of the turbulence.

\section{Expression for the large-scale CRA}
\label{CRA}

The alignment of the CRA with local field lines within $\sim 10$\,pc
is compatible with the widely held view that CRs diffuse
preferentially along field lines. Furthermore, because the power in
the fluctuations on which TeV--PeV CRs scatter ($\sim
10^{-4}-1$\,pc\,$\ll 10$\,pc) is small with respect to that in the
ordered field \cite{Schwadron2014}, we assume pitch-angle diffusion of
CRs in a $\sim 10$\,pc-long, uniform flux tube containing the
Earth, denoting by $D_{\mu\mu}$ the corresponding pitch-angle diffusion
coefficient, ($\mu$ is the cosine of the CR pitch-angle, the angle
between the CR momentum and the ordered magnetic field). 
Since we study here the shape of the CRA, and not its absolute
amplitude, we normalize the latter to 1. Assuming homogeneous
turbulence, and that the problem is 1D and stationary, the normalised CRA at
Earth~\cite{Giacinti:2016tld} is
\begin{equation}
g(\mu) = \frac{\int_{0}^{\mu} {\rm d}\mu' \, \left(1-\mu'^{2}\right)/D_{\mu'\mu'}}{\int_{0}^{1} {\rm d}\mu' \, \left(1-\mu'^{2}\right)/D_{\mu'\mu'}} \,,
\label{EqnCRA}
\end{equation}
provided that the distance between the boundaries of the flux tube and
the Earth is large enough for the diffusion approximation to apply. A
delimitation of the parameter space in which this condition is
satisfied is given in~\cite{Giacinti:2016tld}. Outside this range, the
CRA is determined by unknown, external boundary conditions. From now
on, we only consider cases where the CRA is proportional to
$g(\mu)$. In the following, we use the (dimensionless) pitch-angle
scattering frequency $\nu(\mu)=2D_{\mu\mu}/(1-\mu^{2})\times (l/v)$,
where $l$ denotes the outer scale of the turbulence and $v\approx c$
is the CR velocity.  The physically unmotivated case of isotropic
pitch-angle scattering corresponds $\nu(\mu)=\,$constant, and to
a purely dipole anisotropy, $g(\mu)=\mu$.

$D_{\mu\mu}$ (cf. Eq.~(20) in Ref.~\cite{Giacinti:2016tld}) can be
expressed in terms of a phenomenological resonance function,
$\mathcal{R}_{n}(k_\parallel v_\parallel - \omega + n \Omega)$, where
$k_{\parallel}$ denotes the parallel component of the wavevector {\bf
  k}, $\omega$ the angular frequency of waves, $\Omega$ the CR
gyrofrequency, and $v_{\parallel} = v \mu$. We use broad
($\mathcal{R}_{n}^{\rm B}$) and narrow ($\mathcal{R}_{n}^{\rm N}$)
resonance functions, taken respectively from~\cite{Yan:2007uc}
and~\cite{Chandran:2000hp}:
\begin{displaymath}
  \mathcal{R}_{n}^{\rm B} = \frac{\sqrt{\pi}}{\left|k_{\parallel}\right| v_{\perp} \delta\mathcal{M}_{\rm A}^{1/2}} 
\exp \left( - \frac{(k_\parallel v_\parallel - \omega + n \Omega)^{2}}{k_{\parallel}^{2} v_{\perp}^{2} \delta\mathcal{M}_{\rm A}}\right)
\end{displaymath}
\begin{equation}
  \mathcal{R}_{n}^{\rm N} = \frac{\tau^{-1}}{(k_\parallel v_\parallel - \omega + n \Omega)^{2} + \tau^{-2}} \;,
\label{Eqns_Rn}
\end{equation}
where $v_{\perp} = v \sqrt{1-\mu^{2}}$. $\mathcal{R}_{n}^{\rm B}$
takes into account the broadening of the resonance due to fluctuations
of the magnetic field strength, expressed as the parameter
$\delta\mathcal{M}_{\rm A}\lesssim 1$, whose local value is
poorly constrained. $\mathcal{R}_{n}^{\rm N}$ assumes, instead, that the broadening
of the resonance, described by $\tau$, is dominated by the Lagrangian
correlation time of the turbulence.

\section{Results}
\label{Results}

We use here models with $D_{\mu\mu}(-\mu)=D_{\mu\mu}(\mu)$. Thus, $\nu(-\mu)=\nu(\mu)$ and $g(-\mu)=-g(\mu)$, and we plot $\nu$ and $g$ on $0 \leq \mu \leq 1$ only. Typically, $l \sim (1-100)$\,pc, so the dimensionless CR rigidity $\epsilon = v/l\Omega$ is $\sim 10^{-4}-10^{-1}$ for $\sim {\rm PeV}$ CRs. We assume that the field points in the direction $(l,b)=(47^{\circ},25^{\circ})$, which is compatible with~\cite{Frisch:2012zj}.


\begin{figure*}
  \centerline{\includegraphics[width=0.33\textwidth]{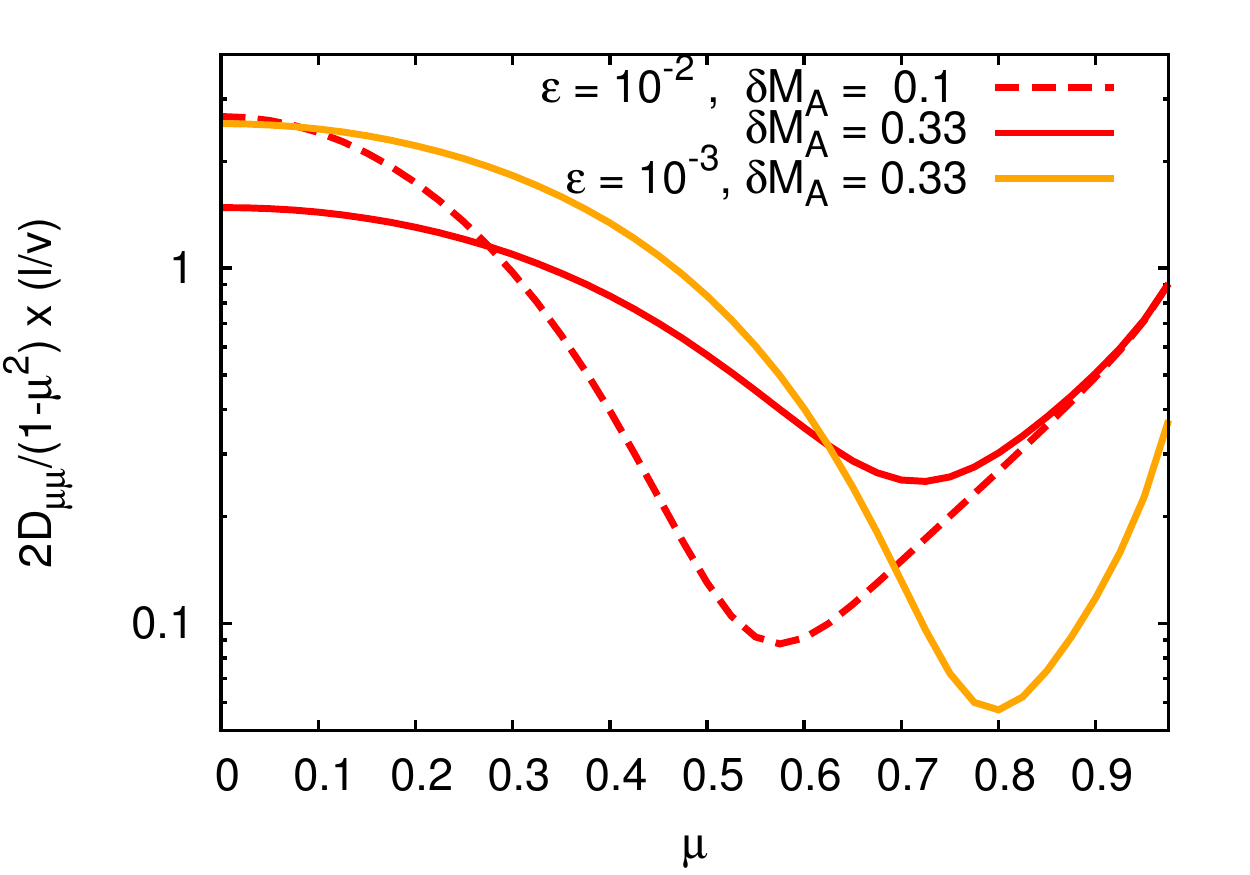}
              \hfil
              \includegraphics[width=0.33\textwidth]{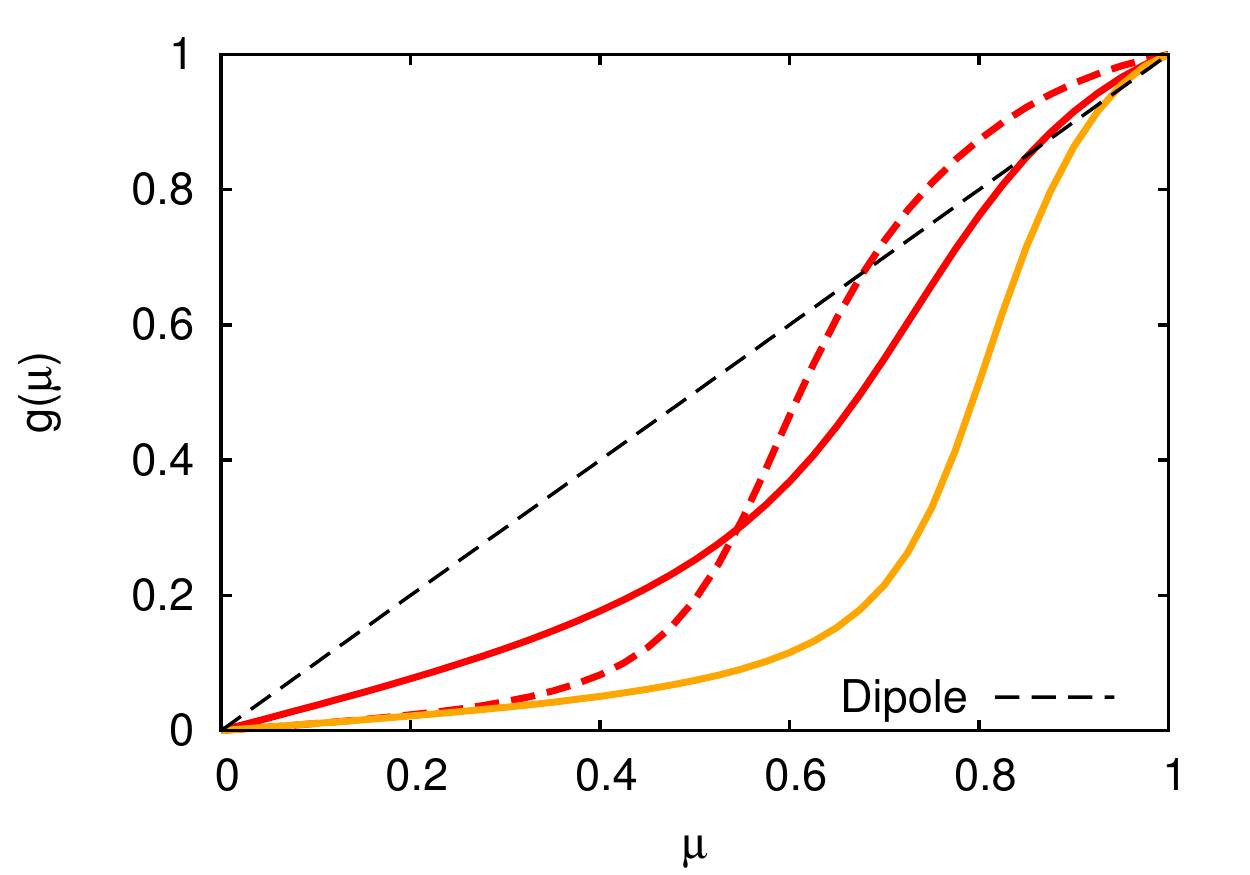}
              \hfil
              \includegraphics[width=0.33\textwidth]{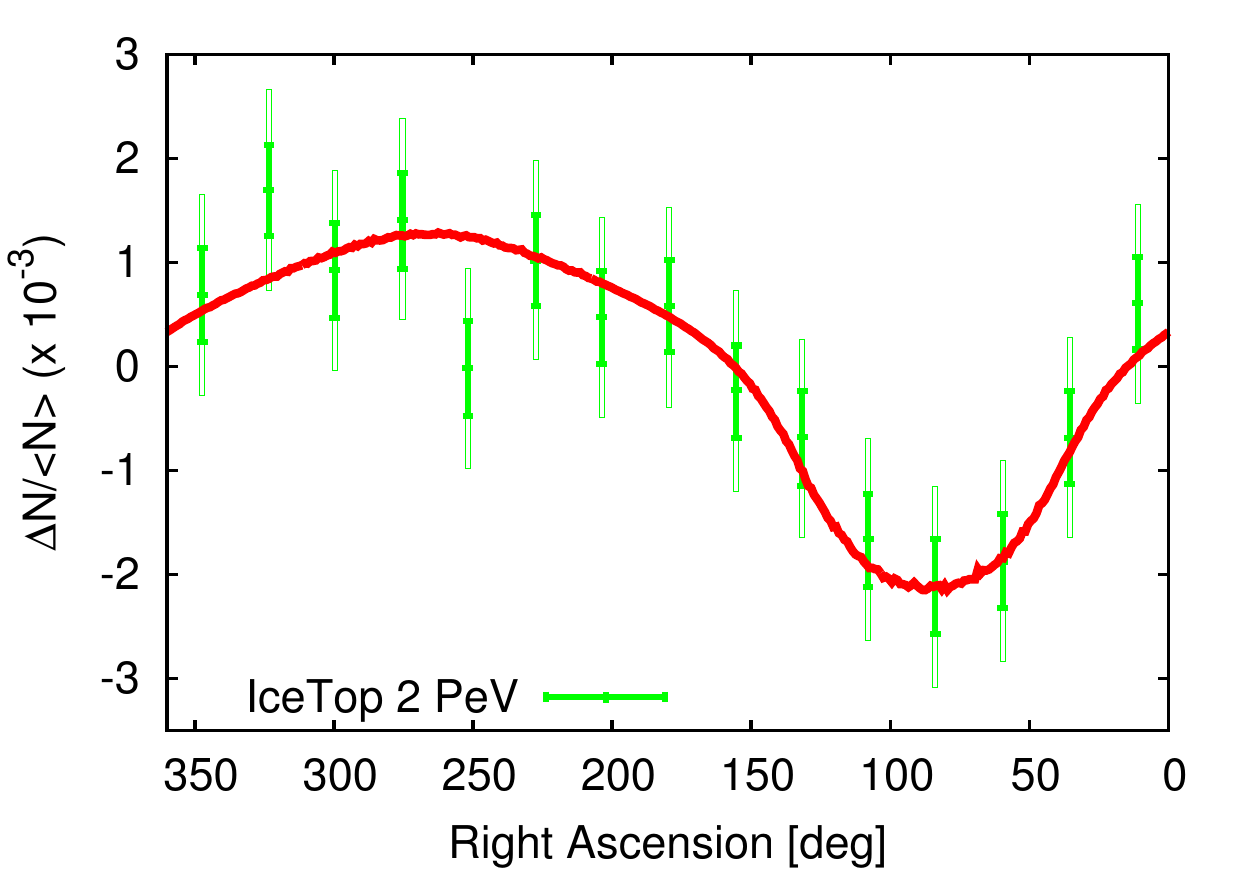}
              }
  \centerline{\includegraphics[width=0.33\textwidth]{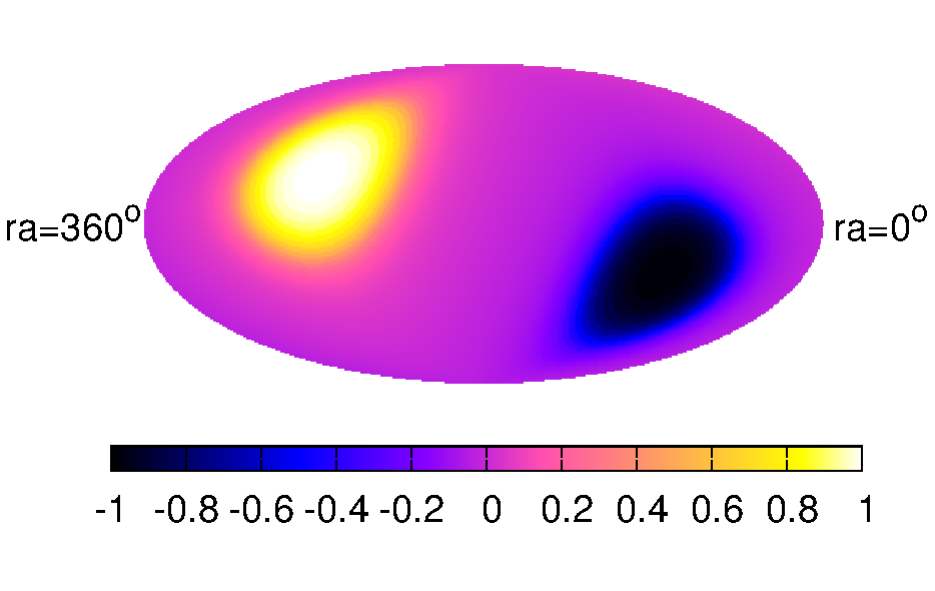}
              \hfil
              \includegraphics[width=0.33\textwidth]{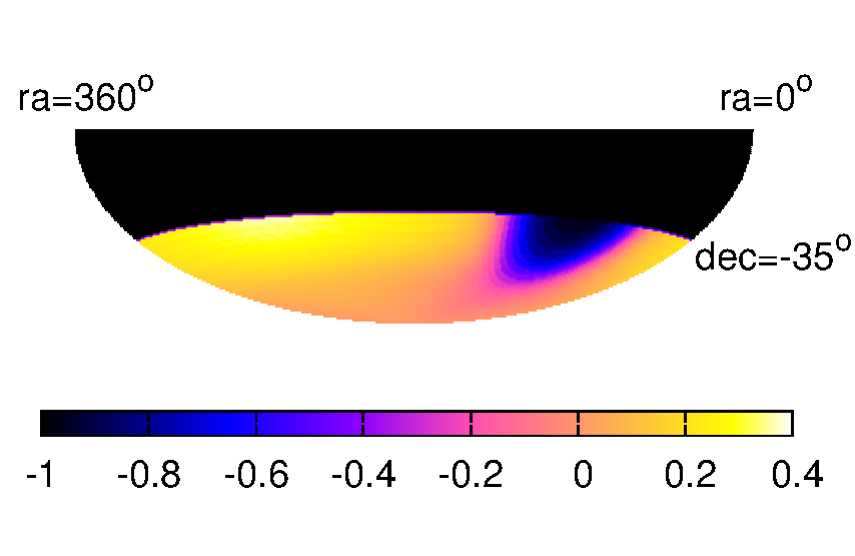}
              \hfil
              \includegraphics[width=0.33\textwidth]{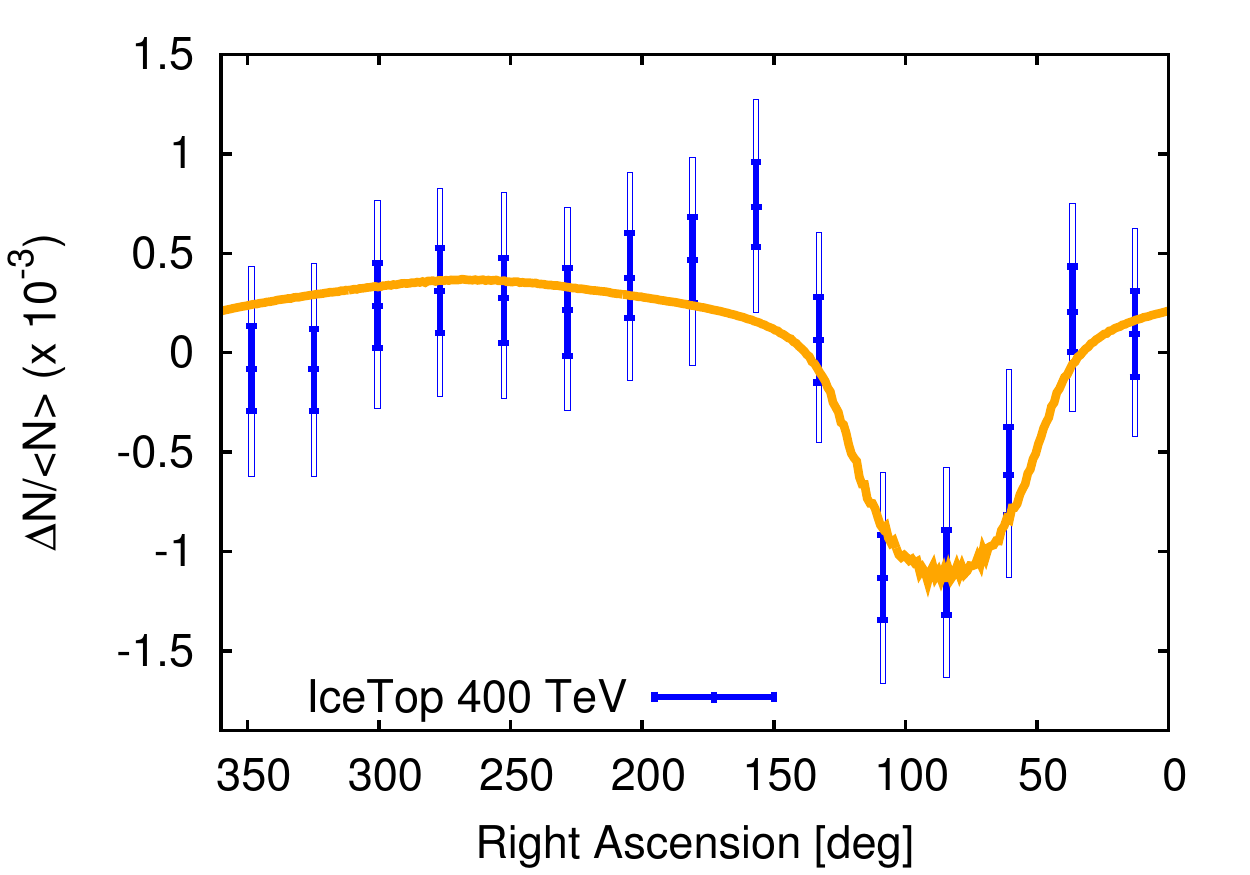}
              }
\caption{GS turbulence with $\mathcal{R}_{n}^{\rm B}$: $\nu(\mu)$ ({\it upper left panel}), $g(\mu)$ ({\it upper centre}), and relative CR intensity at $-75^{\circ} \leq {\rm decl.} \leq -35^{\circ}$ versus R.A., compared with the 2\,PeV (resp. 400\,TeV) data from IceTop ({\it upper right, resp. lower right}). See upper left panel for the parameters of each set of lines. {\it Lower left panel:} Anisotropy $g$ in equatorial coordinates for $\epsilon = 10^{-3}$ and $\delta \mathcal{M}_{\rm A} = 0.33$. {\it Lower centre panel:} CRA in the field of view of IceTop (calculated with respect to the average flux in each declination band) for the same $\epsilon$ and $\delta \mathcal{M}_{\rm A}$, and with minimum amplitude renormalized to $-1$.}
\label{GSturb}
\end{figure*}

First, we consider Goldreich-Sridhar (GS) turbulence~\cite{Goldreich:1994zz}, with a spectrum of Alfv\'en and pseudo-Alfv\'en waves following the prescription of~\cite{Cho:2001hf}: $\mathcal{I}_{\rm A,S}({\bf k}) \propto k_\perp^{-10/3} \exp (-k_{\parallel}l^{1/3}/k_\perp^{2/3})$. In Ref.~\cite{Giacinti:2016tld}, we also studied spectra with an abrupt cutoff in $k_{\parallel}$ and found qualitatively similar trends. The resonance function $\mathcal{R}_{n}^{\rm N}$ provides so little scattering that the CRA is given by $g(\mu)$ only in a small fraction of parameter space, in which it anyway does not fit IceTop data. For GS turbulence, we then only use $\mathcal{R}_{n}^{\rm B}$. We calculate numerically $D_{\mu\mu}$, and plot $\nu(\mu)$ (resp. the CRA, $g(\mu)$) in the upper left (resp. centre) panel of Fig.~\ref{GSturb}, for 3 sets of values of $\{\epsilon,\delta \mathcal{M}_{\rm A}\}$ that fit well IceTop data~\cite{Aartsen:2012ma}: $\{10^{-2},0.1\}$ (red dashed line), $\{10^{-2},0.33\}$ (red solid), $\{10^{-3},0.33\}$ (orange solid). For these parameters, $\nu$ exhibits a minimum in the range $|\mu|\simeq 0.55-0.8$ which corresponds to the transition between two regions: At smaller $|\mu|$ (i.e. perpendicular to field lines), scattering is dominated by the $n=0$ contribution of pseudo-Alfv\'en modes, whereas at larger $|\mu|$, Alfv\'en modes dominate. At fixed $\epsilon$, and when $\delta \mathcal{M}_{\rm A}$ increases, the minimum of $\nu$ occurs at larger $|\mu|$, cf. the two red lines. Indeed, the width of the bump around $\mu=0$ increases for broader resonances. The minimum of $\nu$ corresponds to the sharp increase of $g$ at large $\mu$, see centre panel and Eq.~(\ref{EqnCRA}). This leads to excesses/deficits in the CRA along field lines ($\mu=\pm 1$) that are narrower than for a dipole (black dashed line). We plot the anisotropy in (R.A.,\,decl.) for $\{10^{-3},0.33\}$ in the lower left panel. Tight cold/hot spots are present along the field, and a rather wide flat region lies in-between (magenta). This leads to a Southern sky map similar to what IceTop observes (lower centre panel): A narrow cold spot (dark blue) surrounded with a flat CR intensity. The two right panels show our predicted relative intensities at $-75^{\circ} \leq {\rm decl.} \leq -35^{\circ}$ versus IceTop data sets for a {\em fixed} set of turbulence parameters ($\delta \mathcal{M}_{\rm A}=0.33$): At low CR energy ($\epsilon=10^{-3}$), we can fit well the 400\,TeV data, and by increasing the energy by a factor 10 ($\epsilon=10^{-2}$) ---comparable to the factor 5 in the data, the 2\,PeV data is also well fitted. We note that the existing data can also exclude a non-negligible range of possible parameter values, see~\cite{Giacinti:2016tld} for a wider scan.

\begin{figure*}
  \centerline{\includegraphics[width=0.33\textwidth]{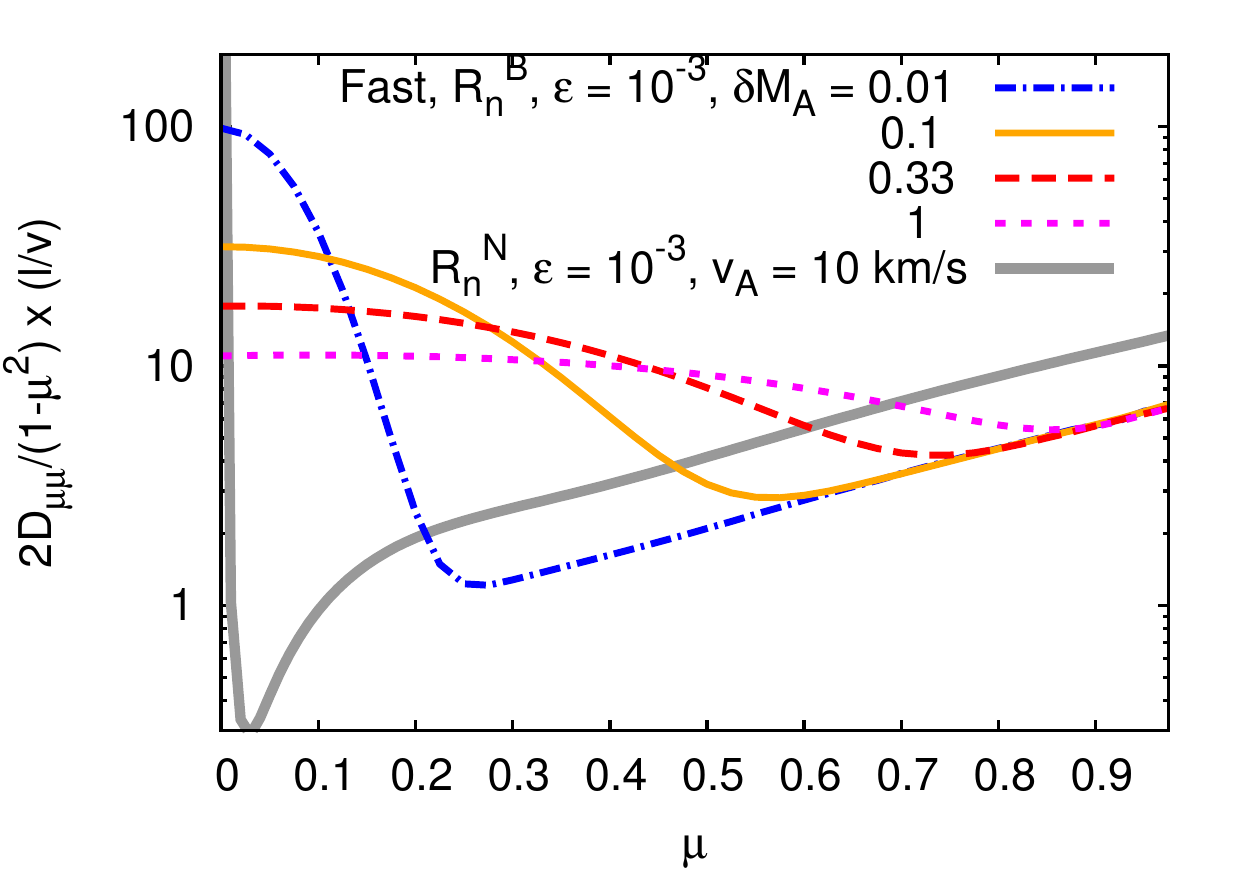}
              \hfil
              \includegraphics[width=0.33\textwidth]{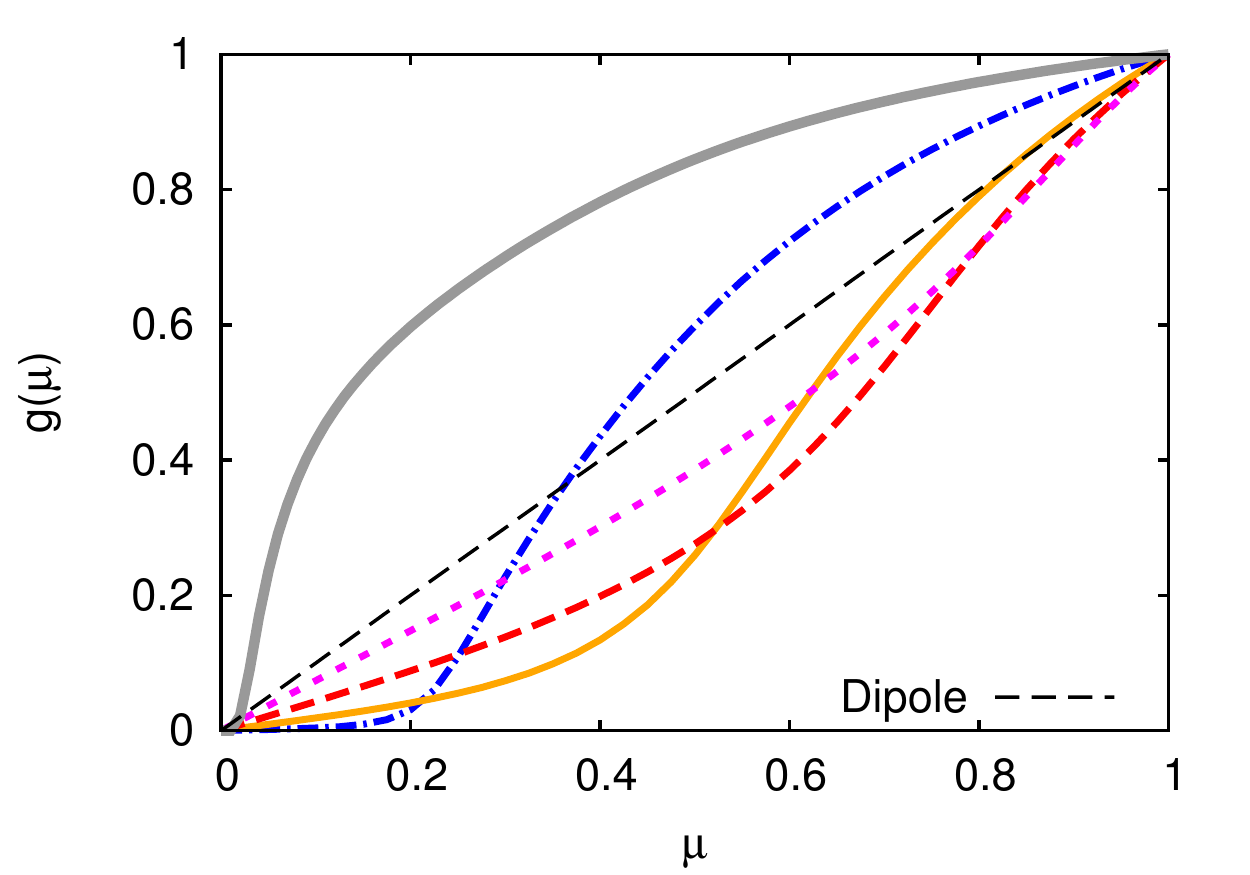}
              \hfil
              \includegraphics[width=0.33\textwidth]{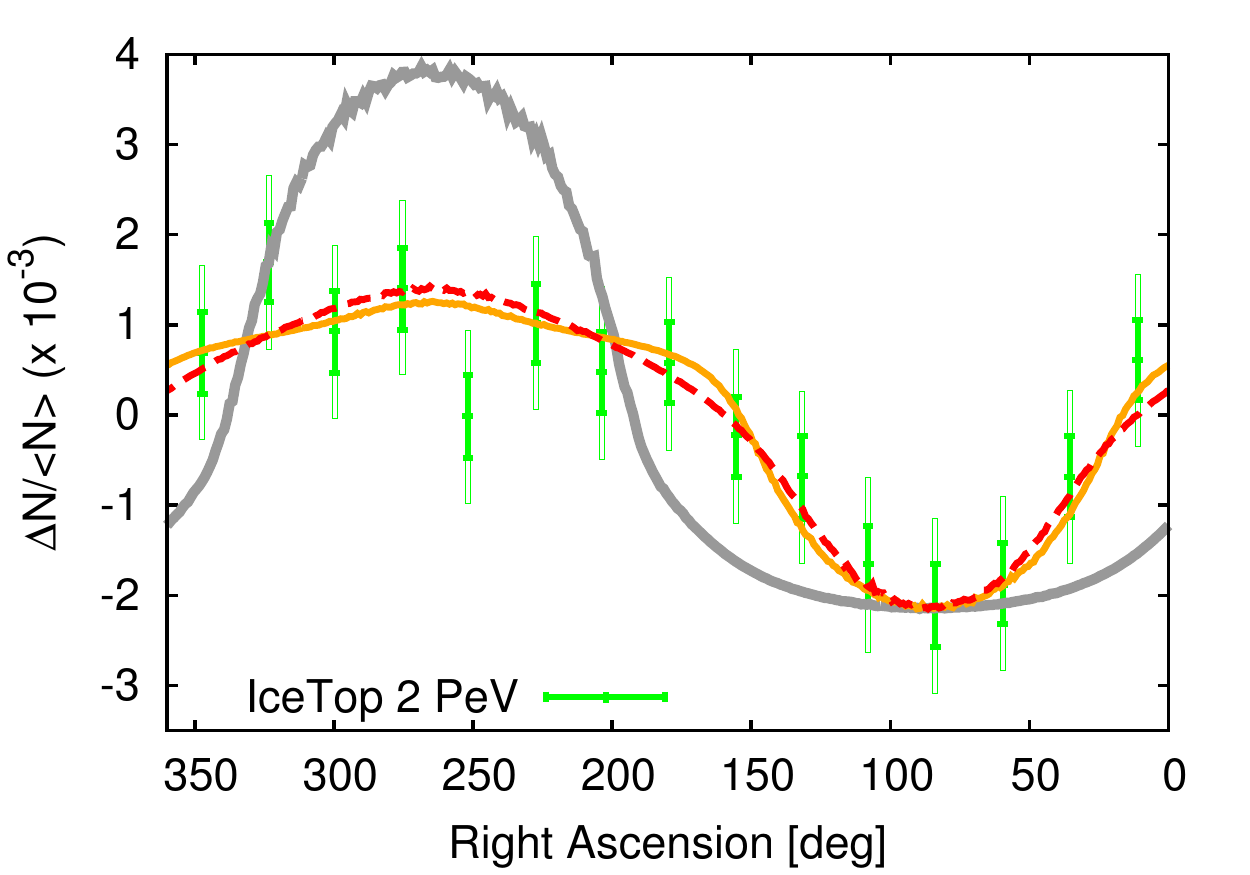}
              }
  \caption{Isotropic fast modes: $\nu(\mu)$ ({\it left panel}), $g(\mu)$ ({\it center panel}), and relative CR intensity at $-75^{\circ} \leq {\rm decl.} \leq -35^{\circ}$ versus R.A., compared with the 2\,PeV data from IceTop ({\it right panel}). See left panel for the parameters of each set of lines.}
\label{FastModes}
\end{figure*}

Second, we study isotropic fast mode turbulence with a power spectrum $\mathcal{I}_{\rm F}(k) \propto k^{-3/2}$, as suggested by~\cite{ChoLazarian2002}. In Fig.~\ref{FastModes}, we plot $\nu$ (left), $g$ (centre), and the relative CR intensity at $-75^{\circ} \leq {\rm decl.} \leq -35^{\circ}$ versus IceTop 2\,PeV data (right). We consider $\mathcal{R}_{n}^{\rm B}$ with $\delta \mathcal{M}_{\rm A}=0.01$ (blue dashed-dotted lines), 0.1 (orange solid), 0.33 (red dashed), 1 (magenta dotted), as well as $\mathcal{R}_{n}^{\rm N}$ for a local Alfv\'en velocity $v_{\rm A} = 10$\,km/s (grey solid), and with $\tau = \sqrt{l}/v_{\rm A}\sqrt{k}$. We take $\epsilon = 10^{-3}$ for the plot of $\nu(\mu)$. We find that $g$ does not depend on $\epsilon$ here. As for pseudo-Alfv\'en modes above, the $n=0$ term for fast modes is now responsible for the peak of $\nu$ around $\mu=0$. This peak grows in width when $\delta \mathcal{M}_{\rm A}$ increases. With the narrow resonance function, it becomes very sharp and the minimum of $\nu$ is close to $\mu=0$, see grey line. This results in a qualitatively different shape of the CRA, cf. middle panel: Very wide cold/hot spots with a rather flat intensity inside. They take nearly half of the sky each ($\simeq 80^{\circ}$ half-width). IceTop data clearly rules out this {\em a priori} acceptable scenario, cf. right panel. For $\mathcal{R}_{n}^{\rm B}$, the 2\,PeV data is well fitted with $\delta \mathcal{M}_{\rm A}=0.1$ and 0.33, where the deficit at R.A.~$\approx 90^{\circ}$ reaches its minimum width. However, the 400\,TeV data is not well fitted. Decreasing $\delta \mathcal{M}_{\rm A}$ makes the deficit too wide (blue lines), while increasing $\delta \mathcal{M}_{\rm A}$ to 1 (magenta lines) makes the anisotropy tend towards a dipole, which again increases the size of the deficit.

\section{Discussion and perspectives}
\label{Discussion}

In all models above, the CRA exhibits a flattening in directions
perpendicular to field lines, due to the peak of the scattering
frequency around $\mu=0$. Interestingly, this signature is compatible
with the observational data. Moderately broad resonance functions are
able to produce deficits/excesses along the field direction ($\mu=\pm
1$) that are narrower than those of a dipole. Too broad a resonance
would, however, result in an approximately dipole anisotropy with
constant $\nu$, see the magenta lines in Fig.~\ref{FastModes}. On the
other hand, narrow resonance functions are disfavoured by IceTop
data. Nonetheless, we cannot formally exclude narrow resonance
functions for models of turbulence in which $\nu$ is so small that the
CR mean free path is $\gg 100$\,pc, such as in~\cite{Chandran:2000hp},
since our study does not then apply. A very low CR scattering rate
locally is not impossible, but shifts the problem of CR confinement
and the explanation of its anisotropy to large distance. The increase
with CR energy of the width of the deficit between the two IceTop data
sets may hint at a $|{\bf k}|$-dependent anisotropy in the turbulence
power-spectrum, such as in GS turbulence. Other possibilities, such as
an energy-dependence of the resonance function should also be
explored. As for fast modes, they should suffer from
damping~\cite{Yan:2007uc}, which we did not take into account, and
which may have an effect on the CRA. A combined analysis of all
available CRA data
should yield tighter constraints than those presented in Sect.~\ref{Results}. Distortions of the CRA by heliospheric fields 
may complicate the picture at $\lesssim 10$\,TeV. The recent data from
the Tibet Air Shower Array~\cite{2017arXiv170107144T} hints at the
presence of a narrow hot spot in the Northern hemisphere, in their
300\,TeV data set (cf. their Fig.~4), as would be expected within our
model and for a $D_{\mu\mu}$ symmetric with respect to $\mu=0$.

\section{Conclusions}
\label{Conclusions}

Assuming pitch-angle diffusion of Galactic CRs in our local
environment, we deduced the shape of the LS CRA, see
Eq.~(\ref{EqnCRA}). In general, it is not a pure dipole, but contains
imprints of the still poorly known properties of $(i)$ the local
interstellar turbulent magnetic fields (e.g. power-spectrum), and
$(ii)$ CR transport (via $\mathcal{R}_{n}$). We find that the existing
observational data already puts constraints on these. A {\em
  moderately} broad resonance function seems to be favoured. IceTop
2\,PeV data can be fitted either with GS turbulence or isotropic fast
modes (see parameters in Figs.~\ref{GSturb} and~\ref{FastModes}), but
only the former can reproduce the change in the width of the deficit
between the 400\,TeV and 2\,PeV data sets. We suggest that the shape
of the LS CRA can be used as a new observable.

\nocite{*}
\bibliographystyle{elsarticle-num}
\bibliography{references2}

\begin{thebibliography}{10}
\expandafter\ifx\csname url\endcsname\relax
  \def\url#1{\texttt{#1}}\fi
\expandafter\ifx\csname urlprefix\endcsname\relax\def\urlprefix{URL }\fi
\expandafter\ifx\csname href\endcsname\relax
  \def\href#1#2{#2} \def\path#1{#1}\fi

\bibitem{Giacinti:2016tld}
G.~Giacinti, J.~G. Kirk, {Large-Scale Cosmic-Ray Anisotropy as a Probe of
  Interstellar Turbulence}, ApJ 835 (2017) 258.
\newblock \href {http://arxiv.org/abs/1610.06134} {\path{arXiv:1610.06134}},
  \href {http://dx.doi.org/10.3847/1538-4357/835/2/258}
  {\path{doi:10.3847/1538-4357/835/2/258}}.

\bibitem{Ahlers:2016njd}
M.~{Ahlers}, {Deciphering the Dipole Anisotropy of Galactic Cosmic Rays},
  Physical Review Letters 117~(15) (2016) 151103.
\newblock \href {http://arxiv.org/abs/1605.06446} {\path{arXiv:1605.06446}},
  \href {http://dx.doi.org/10.1103/PhysRevLett.117.151103}
  {\path{doi:10.1103/PhysRevLett.117.151103}}.

\bibitem{Schwadron2014}
N.~A. {Schwadron}, F.~C. {Adams}, E.~R. {Christian}, P.~{Desiati}, P.~{Frisch},
  H.~O. {Funsten}, J.~R. {Jokipii}, D.~J. {McComas}, E.~{Moebius}, G.~P.
  {Zank}, {Global Anisotropies in TeV Cosmic Rays Related to the Sun's Local
  Galactic Environment from IBEX}, Science 343 (2014) 988--990.
\newblock \href {http://dx.doi.org/10.1126/science.1245026}
  {\path{doi:10.1126/science.1245026}}.

\bibitem{Frisch:2012zj}
P.~C. {Frisch}, B.-G. {Andersson}, A.~{Berdyugin}, V.~{Piirola},
  R.~{DeMajistre}, H.~O. {Funsten}, A.~M. {Magalhaes}, D.~B. {Seriacopi}, D.~J.
  {McComas}, N.~A. {Schwadron}, J.~D. {Slavin}, S.~J. {Wiktorowicz}, {The
  Interstellar Magnetic Field Close to the Sun. II.}, ApJ 760 (2012) 106.
\newblock \href {http://arxiv.org/abs/1206.1273} {\path{arXiv:1206.1273}},
  \href {http://dx.doi.org/10.1088/0004-637X/760/2/106}
  {\path{doi:10.1088/0004-637X/760/2/106}}.

\bibitem{Frisch:2015hfa}
P.~C. {Frisch}, A.~{Berdyugin}, V.~{Piirola}, A.~M. {Magalhaes}, D.~B.
  {Seriacopi}, S.~J. {Wiktorowicz}, B.-G. {Andersson}, H.~O. {Funsten}, D.~J.
  {McComas}, N.~A. {Schwadron}, J.~D. {Slavin}, A.~J. {Hanson}, C.-W. {Fu},
  {Charting the Interstellar Magnetic Field causing the Interstellar Boundary
  Explorer (IBEX) Ribbon of Energetic Neutral Atoms}, ApJ 814 (2015) 112.
\newblock \href {http://arxiv.org/abs/1510.04679} {\path{arXiv:1510.04679}},
  \href {http://dx.doi.org/10.1088/0004-637X/814/2/112}
  {\path{doi:10.1088/0004-637X/814/2/112}}.

\bibitem{Aartsen:2012ma}
M.~G. {Aartsen}, R.~{Abbasi}, Y.~{Abdou}, M.~{Ackermann}, J.~{Adams}, J.~A.
  {Aguilar}, M.~{Ahlers}, D.~{Altmann}, K.~{Andeen}, J.~{Auffenberg}, et~al.,
  {Observation of Cosmic-Ray Anisotropy with the IceTop Air Shower Array}, ApJ
  765 (2013) 55.
\newblock \href {http://arxiv.org/abs/1210.5278} {\path{arXiv:1210.5278}},
  \href {http://dx.doi.org/10.1088/0004-637X/765/1/55}
  {\path{doi:10.1088/0004-637X/765/1/55}}.

\bibitem{Zhang:2014dsu}
M.~{Zhang}, P.~{Zuo}, N.~{Pogorelov}, {Heliospheric Influence on the Anisotropy
  of TeV Cosmic Rays}, ApJ 790 (2014) 5.
\newblock \href {http://dx.doi.org/10.1088/0004-637X/790/1/5}
  {\path{doi:10.1088/0004-637X/790/1/5}}.

\bibitem{Yan:2007uc}
H.~{Yan}, A.~{Lazarian}, {Cosmic-Ray Propagation: Nonlinear Diffusion Parallel
  and Perpendicular to Mean Magnetic Field}, ApJ 673 (2008) 942--953.
\newblock \href {http://arxiv.org/abs/0710.2617} {\path{arXiv:0710.2617}},
  \href {http://dx.doi.org/10.1086/524771} {\path{doi:10.1086/524771}}.

\bibitem{Chandran:2000hp}
B.~D.~G. {Chandran}, {Scattering of Energetic Particles by Anisotropic
  Magnetohydrodynamic Turbulence with a Goldreich-Sridhar Power Spectrum},
  Physical Review Letters 85 (2000) 4656--4659.
\newblock \href {http://arxiv.org/abs/astro-ph/0008498}
  {\path{arXiv:astro-ph/0008498}}, \href
  {http://dx.doi.org/10.1103/PhysRevLett.85.4656}
  {\path{doi:10.1103/PhysRevLett.85.4656}}.

\bibitem{Goldreich:1994zz}
P.~{Goldreich}, S.~{Sridhar}, {Toward a theory of interstellar turbulence. 2:
  Strong alfvenic turbulence}, ApJ 438 (1995) 763--775.
\newblock \href {http://dx.doi.org/10.1086/175121} {\path{doi:10.1086/175121}}.

\bibitem{Cho:2001hf}
J.~{Cho}, A.~{Lazarian}, E.~T. {Vishniac}, {Simulations of Magnetohydrodynamic
  Turbulence in a Strongly Magnetized Medium}, ApJ 564 (2002) 291--301.
\newblock \href {http://arxiv.org/abs/astro-ph/0105235}
  {\path{arXiv:astro-ph/0105235}}, \href {http://dx.doi.org/10.1086/324186}
  {\path{doi:10.1086/324186}}.

\bibitem{ChoLazarian2002}
J.~{Cho}, A.~{Lazarian}, {Compressible Sub-Alfv{\'e}nic MHD Turbulence in Low-
  {$\beta$} Plasmas}, Physical Review Letters 88~(24) (2002) 245001.
\newblock \href {http://arxiv.org/abs/astro-ph/0205282}
  {\path{arXiv:astro-ph/0205282}}, \href
  {http://dx.doi.org/10.1103/PhysRevLett.88.245001}
  {\path{doi:10.1103/PhysRevLett.88.245001}}.

\bibitem{2017arXiv170107144T}
{Tibet AS-gamma Collaboration}, M.~{Amenomori}, {Northern sky Galactic Cosmic
  Ray anisotropy between 10-1000 TeV with the Tibet Air Shower Array}, ArXiv
  e-prints. \href {http://arxiv.org/abs/1701.07144} {\path{arXiv:1701.07144}}.

\end{thebibliography}

\end{document}